\documentclass[aps,prb,superscriptaddress,twocolumn,showpacs,preprintnumbers,amsmath,amssymb]{revtex4}

\usepackage[dvips]{graphicx,color}
\usepackage{dcolumn}
\usepackage{bm}
\usepackage{ulem}
\usepackage{soul}
\begin{document}

\title{Indication of intrinsic spin Hall effect in 4d and 5d transition metals}

\author{M. Morota}
\affiliation{Institute for Solid State Physics, University of Tokyo, 5-1-5 Kashiwa-no-ha, Kashiwa, Chiba 277-8581, Japan}
\author{Y. Niimi}
\email{niimi@issp.u-tokyo.ac.jp}
\affiliation{Institute for Solid State Physics, University of Tokyo, 5-1-5 Kashiwa-no-ha, Kashiwa, Chiba 277-8581, Japan}
\author{K. Ohnishi}
\affiliation{Institute for Solid State Physics, University of Tokyo, 5-1-5 Kashiwa-no-ha, Kashiwa, Chiba 277-8581, Japan}
\author{D. H. Wei}
\affiliation{Institute for Solid State Physics, University of Tokyo, 5-1-5 Kashiwa-no-ha, Kashiwa, Chiba 277-8581, Japan}
\author{T. Tanaka}
\affiliation{Department of Physics, Nagoya Universtiy, Furo-cho, Nagoya 464-8602, Japan}
\author{H. Kontani}
\affiliation{Department of Physics, Nagoya Universtiy, Furo-cho, Nagoya 464-8602, Japan}
\author{T. Kimura}
\altaffiliation{Present address: Inamori Frontier Research Center, Kyushu University, 744 Motooka, Nishi-ku, Fukuoka 819-0395, Japan}
\affiliation{Institute for Solid State Physics, University of Tokyo, 5-1-5 Kashiwa-no-ha, Kashiwa, Chiba 277-8581, Japan}
\author{Y. Otani}
\email{yotani@issp.u-tokyo.ac.jp}
\affiliation{Institute for Solid State Physics, University of Tokyo, 5-1-5 Kashiwa-no-ha, Kashiwa, Chiba 277-8581, Japan}
\affiliation{RIKEN-ASI, 2-1 Hirosawa, Wako, Saitama 351-0198, Japan}

\date{\today}

\begin{abstract}
We have investigated spin Hall effects in 4$d$ and 5$d$ transition metals, 
Nb, Ta, Mo, Pd and Pt, by incorporating the spin absorption method 
in the lateral spin valve structure; 
where large spin current preferably relaxes into the transition metals, 
exhibiting strong spin-orbit interactions. 
Thereby nonlocal spin valve measurements enable us to evaluate 
their spin Hall conductivities. 
The sign of the spin Hall conductivity changes systematically 
depending on the number of $d$ electrons. 
This tendency is in good agreement with the recent theoretical 
calculation based on the intrinsic spin Hall effect. 
\end{abstract}


\maketitle

\section{Introduction}

Spin current, a flow of the spin angular momentum, 
is an important physical quantity 
to operate spintronic devices.~\cite{Maekawa} 
The spin Hall effect (SHE) is widely recognized as a phenomenon 
that converts charge current to the spin current 
requiring neither external magnetic fields 
nor ferromagnets.~\cite{SHE1,SHE2,SHE3} 
The search for materials exhibiting large SHEs is 
therefore a prime task for further advancement of spintronic devices. 
Since the SHE originates from spin-dependent scattering events, 
materials with large spin-orbit interactions can be good candidates 
for efficient generation of the spin current.
It is, however, difficult to study the SHEs in such materials since 
the spin diffusion length is extremely short 
(of the order of several nanometers).
We have established the sensitive electrical detection technique of the SHE 
using the spin current absorption effect.~\cite{Kimura_SHE,Vila_SHE}  
The greatest advantage of this technique is that
one can measure the SHE as well as the spin diffusion length of materials 
with large spin-orbit interactions on the same device.   
This enables us to obtain the SH conductivity as well as the SH angle 
which is defined as the ratio of SH conductivity and charge conductivity.  

Spin-dependent Hall effects have been theoretically discussed 
in terms of two distinct physical mechanisms.  
One is the extrinsic mechanism induced by the impurity 
scattering~\cite{Extrinsic1,Extrinsic2} 
that was intensively investigated a few decades ago 
as an origin of the anomalous Hall effect (AHE).~\cite{Extrinsic3}
The other is the intrinsic mechanism based on the band-structure effect 
as a manifestation of the Berry phase.~\cite{Intrinsic1,Intrinsic2} 
It was believed that the intrinsic mechanism is limited only in 
very clean systems such as semiconductors 
with high electron mobility.~\cite{murakami,sinova}
Recently, intrinsic AHEs where spin-orbit 
interaction together with the interband mixing results in 
an intrinsic anomalous velocity in the transverse direction
have been observed in many systems 
even at room temperature.~\cite{Lee,Miyasako,Jin,AHE1,note_intrinsic} 
In the case of SHEs in nonmagnetic materials, on the other hand,
there is no Hall voltage since the number of spin-up and down electrons 
are exactly same. 
However, when the \textit{pure} spin current 
defined as the difference between 
the spin-up and down currents is injected into such materials, 
both spin-up and spin-down electrons are scattered to the same side, 
which can be detected as a Hall voltage.
Interestingly, recent theoretical studies show that 
the magnitude and sign of SH conductivities 
due to the intrinsic SHE in 4$d$ and 5$d$ transition metals (TMs) 
change systematically in response to the number 
of $d$ electrons.~\cite{Guo,Kontani1,Kontani2,Kontani3} 
Therefore, systematic experiments of the SHEs in such TMs 
should help to find the dominant mechanism of the observed SHE. 

As described above, the most difficult point for transport measurements of 
the SHEs in 4$d$ and 5$d$ TMs is their short spin diffusion lengths. 
The spin absorption technique 
which is detalied in Refs.~\onlinecite{Kimura_SHE} 
and~\onlinecite{Vila_SHE} enables us to perform quantitative 
and systematic studies of the SHEs even in materials with 
short spin diffusion lengths. 
In this paper, we report on measurements of the SHEs 
in various 4$d$ and 5$d$ TMs using the spin absorption technique. 
The experimentally observed SH conductivities of those TMs are 
semiqunatitatively consistent with the recent calculations based 
on the intrinsic SHE. 
This fact strongly supports that the instrinsic machanism of the SHEs 
in 4$d$ and 5$d$ TMs is more dominant than the extrinsic ones.

\begin{figure}
\begin{center}
\includegraphics[width=6.5cm]{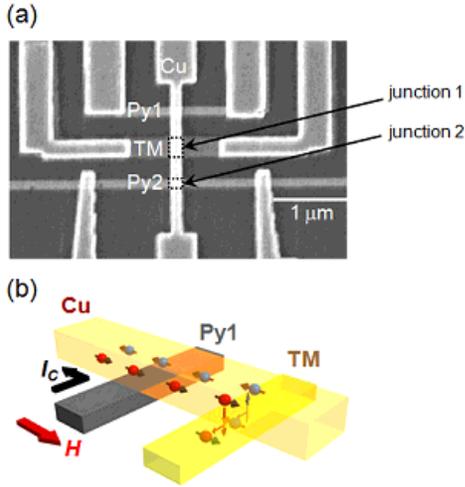}
\caption{(Color online) (a) Scanning electron micrograph of a typical spin Hall device consisting of two Py wires and a TM wire bridged by a Cu strip. (b) Schematic of the mechanism of ISHE due to the spin absorption effect.} \label{fig1}
\end{center}
\end{figure}

\section{Experimental details}

Our device has been fabricated on a thermally oxidized silicon subtrate 
using electron beam lithography on polymethyl-methacrylate (PMMA) resist 
and subsequent lift-off process. 
The device is based on the lateral spin valve structure~\cite{Vila_SHE} 
where a TM middle wire is inserted in between two Permalloy (Py) wires 
as shown in Fig.~1(a). 
The Py wires are 30~nm in thickness ($t_{\rm Py}$) and 
100~nm in width ($w_{\rm Py}$) 
and have been deposited by means of electron beam evaporation.  
Here one Py wire (Py1) has large pads at the edges 
to induce the difference in the switching field. 
In this work, five different TMs (Nb, Ta, Mo, Pd and Pt) 
have been used as a middle wire.  
Nb, Ta and Mo wires were deposited by magnetron sputtering while  
Pd and Pt wires were grown by electron beam evaporation. 
The Cu strip whose thickness ($t_{\rm Cu}$) is 100~nm and 
whose width ($w_{\rm Cu}$) is 150~nm was fabricated 
by a Joule heating evaporator. 
Prior to Cu evaporation, a careful Ar-ion beam etching was carried out 
for 30 s to clean the surfaces of Py and TM wires and 
to obtain highly transparent Ohmic contacts. 

When the spin-polarized current is injected from Py1 into 
the upper side of the Cu strip, there is no net charge current but 
only a \textit{pure} spin current is induced 
on the bottom side of the Cu strip [see Fig.~1(b)]. 
The induced spin current is divided into two segments 
(TM or the bottom side of Cu strip)
at the junction~1 (TM/Cu junction).
When the spin relaxation of the TM wire is much stronger 
than that of the Cu wire, the induced spin current is preferably 
absorbed into the TM wire.  
This leads to a drastic reduction of the spin accumulation voltage 
in the junction~2 (Py2/Cu junction). 
The flowing direction of the spin current 
in the TM wire is perpendicular to the plane of junction~1 
because of its strong spin-orbit interaction, in other words, 
its short spin diffusion length.~\cite{Kimura_SHE,Vila_SHE} 
Therefore, the charge accumulation due to 
the inverse SHE (ISHE) is induced in the TM wire. 
The measurements have been carried out by using an 
ac lock-in amplifier and a He flow cryostat. 
The magnetic field is applied along the easy and hard axes of Py 
for nonlocal spin valve (NLSV)~\cite{review_otani} 
and ISHE measurements, respectively. 

\begin{figure}
\begin{center}
\includegraphics[width=6cm]{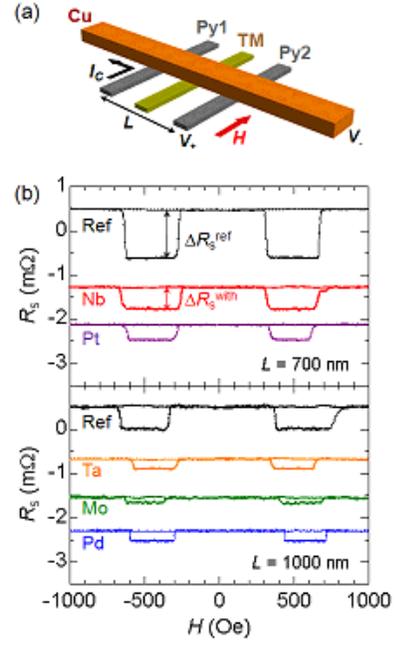}
\caption{(Color online) (a) Schematic of the probe configuration for NLSV measurement. (b) NLSV signals $R_{\rm S}$ with and without TM wires measured at 10~K for $L=700$~nm (upper panel) and $L=1000$~nm (lower panel).} \label{fig2}
\end{center}
\end{figure}

\begin{table*}
\caption{Device dimensions and some characteristic parameters of various TMs.}
\label{table1}
\begin{ruledtabular}
\begin{tabular}{ccccccccc}
Material & $w_{\rm TM}$ (nm) & $t_{\rm TM}$ (nm) & $L$ (nm) & $\eta$ & $\lambda_{\rm TM}$ (nm) & $\sigma_{\rm TM}$ $(10^{3} \Omega^{-1} {\rm cm}^{-1})$ & $\sigma_{\rm SHE}$ $(10^{3} \Omega^{-1} {\rm cm}^{-1})$ & $\alpha_{\rm H} (\%)$ \\ \hline
Nb & 370 & 11 & 700 & $0.35 \pm 0.04$ & $5.9 \pm 0.3$ & 11 & $-(0.10 \pm 0.02)$ & $-(0.87 \pm 0.20)$ \\ 
Ta & 250 & 20 & 1000 & $0.48 \pm 0.04$ & $2.7 \pm 0.4$ & 3 & $-(0.011 \pm 0.003)$ & $-(0.37 \pm 0.11)$ \\ 
Mo & 250 & 20 & 1000 & $0.24 \pm 0.03$ & $8.6 \pm 1.3$ & 28 & $-(0.23 \pm 0.05)$ & $-(0.80 \pm 0.18)$ \\ 
Pd & 250 & 20 & 1000 & $0.37 \pm 0.04$ & $13 \pm 2$ & 22 & $0.27 \pm 0.09$ & $1.2 \pm 0.4$ \\ 
Pt & 100 & 20 & 700 & $0.34 \pm 0.03$ & $11 \pm 2$ & 81 & $1.7 \pm 0.4$ & $2.1 \pm 0.5$ \\ 
\end{tabular}
\end{ruledtabular}
\end{table*}

\section{Results and Discussion}

First, we measure NLSV signals 
to evaluate the spin diffusion lengths of TM wires as well as 
the spin current absorbed into the TM middle wires precisely.~\cite{Vila_SHE} 
As described above, the spin accumulation signal without the TM wires 
$\Delta R_{\rm S}^{\rm ref}$ 
($\equiv \Delta V_{\rm S}^{\rm ref}/I_{\rm C}$, i.e., 
the spin accumulation voltage divided by the charge current)
is reduced to $\Delta R_{\rm S}^{\rm with}$ by inserting the TM middle wires.  
In Fig.~2(b), we show the NLSV signals for various TM insertions. 
All the results exhibit clear spin absorption effects, 
assuring that the spin currents are really absorbed 
into the TM middle wires via the Cu strip. 
From the one-dimensional spin-diffusion model proposed 
by Takahashi and Maekawa,~\cite{Takahashi}  
the normalized spin accumulation signal 
$\Delta R_{\rm S}^{\rm with}/\Delta R_{\rm S}^{\rm ref}$ 
can be calculated as follows:~\cite{notation}
\begin{equation}
\eta \equiv \frac{\Delta R_{\rm S}^{\rm with}}{\Delta R_{\rm S}^{\rm ref}} 
\approx 
\frac{2R_{\rm TM} \sinh (L/\lambda_{\rm Cu})}
{R_{\rm Cu} (\cosh (L/\lambda_{\rm Cu})-1)+ 2R_{\rm TM} \sinh(L/\lambda_{\rm Cu})}. \label{eq1}
\end{equation}
Here $R_{\rm Cu}$ and $R_{\rm TM}$ are the spin resistances 
for Cu and TM, respectively.~\cite{note_spin_resistance} 
The spin resistance for Cu is defined by 
$\frac{\rho_{\rm Cu} \lambda_{\rm Cu}}{w_{\rm Cu}t_{\rm Cu}}$, 
where $\rho_{\rm Cu}$, $\lambda_{\rm Cu}$ are the electrical resistivity and 
the spin diffusion length of Cu.~\cite{Takahashi} 
The spin resistance for TM is defined by 
$\frac{\rho_{\rm TM} \lambda_{\rm TM}}{w_{\rm TM}w_{\rm Cu} \tanh (t_{\rm TM}/\lambda_{\rm TM})}$, 
where $\rho_{\rm TM}$, $\lambda_{\rm TM}$, and $w_{\rm TM}$ are 
the electrical resistivity, the spin diffusion length and the width of 
the TM wire, respectively. 
The hyperbolic tangent term comes from 
the boundary condition where $I_{\rm S} = 0$ at the substrate, 
as detailed in Ref.~\onlinecite{niimi}.
$L$ is the distance between the two Py wires.  
Since $\lambda_{\rm Cu}$ is already known 
from our previous experiments,~\cite{Kimura1} 
$\lambda_{\rm TM}$ can be calculated by using Eq.~(\ref{eq1}). 
The spin diffusion lengths $\lambda_{\rm TM}$ as well as other 
characteristic paprameters for various TM wires 
are summarized in Table~\ref{table1}.  
In the present study $\lambda_{\rm TM}$ is quite short for all the TMs, 
supporting the strong spin-orbit interactions in the TM wires. 

\begin{figure}
\begin{center}
\includegraphics[width=6cm]{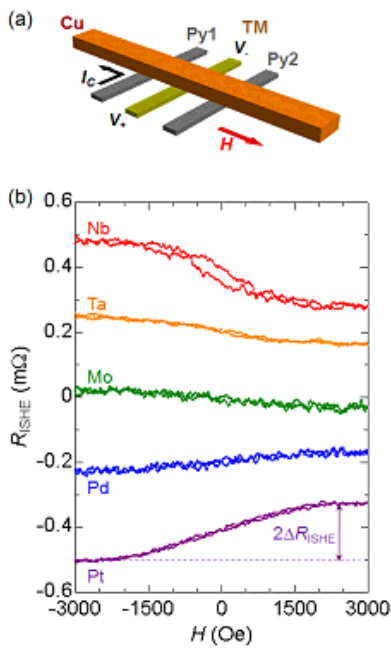}
\caption{(Color online) (a) Schematic of the probe configuration for ISHE measurement. (b) ISHE signals measured at 10~K for various TM wires. The device dimensions are shown in Table~\ref{table1}.
} \label{fig3}
\end{center}
\end{figure}

Next we measure ISHEs for the TM wires. 
Note that the direction of the applied magnetic field in this case
is parallel to the Cu strip corresponding to 
the hard axis of the Py wire as shown in Fig.~3(a). 
In Fig.~3(b) we show the ISHE signals $R_{\rm ISHE}$ measured at 10~K 
for various TM wires. 
For all the TM wires, $R_{\rm ISHE}$ linearly changes 
with the magnetic field below 2000~Oe 
and saturates above 2000~Oe because the magnetization of the Py wire 
fully aligns with the direction of the magnetic field. 
$\Delta R_{\rm ISHE}$ defined in this paper is 
two times smaller than that previously reported by some of 
the present authors.~\cite{Kimura_SHE,Vila_SHE} 
However, we have adapted the current notation 
in order to have consistency with AHE measurements, other SHE measurements 
and theoretical expressions.~\cite{Kontani1,Kontani2,Kontani3}   
Interestingly the sign of the slope below 2000~Oe depends on the TMs; 
the slope is negative for Nb, Ta, and Mo, while it is positive for 
Pd and Pt. 
This clearly shows that the sign of the SH conductivity changes depending 
on the kinds of TMs. 
A similar material dependence of the sign of SH conductivity 
has been reported in Refs.~\onlinecite{Hoffmann1} and~\onlinecite{Hoffmann2}, 
where the spin pumping method has been used to measure ISHEs.

\begin{figure}
\begin{center}
\includegraphics[width=6.5cm]{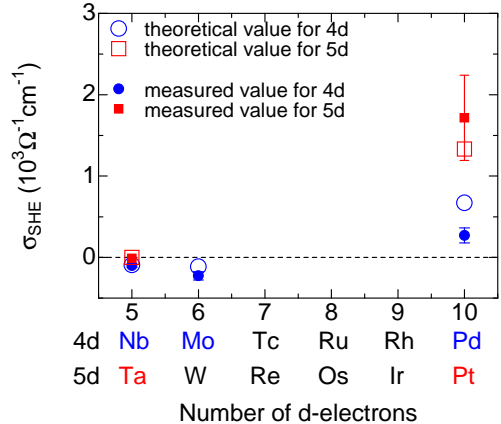}
\caption{(Color online) Experimentally measured (closed symbols) and theoretically calculated (open symbols) spin Hall conductivities as a function of the number of $d$ electrons for 4$d$ (circle) and 5$d$ (square) TMs. } \label{fig4}
\end{center}
\end{figure}

According to the theory on the intrinsic SHE 
in $d$-electron systems,~\cite{Kontani3} 
the SH conductivity in TMs is approximately 
given by the following equation;  
$\sigma_{\rm SHE} \approx (e/4a)\cdot \langle \mbox{\boldmath$l$} \cdot \mbox{\boldmath$s$} \rangle /\hbar^{2}$, where $a$ and 
$\langle \mbox{\boldmath$l$} \cdot \mbox{\boldmath$s$} \rangle$ 
are the lattice constant and the expectation value of
 the LS coupling, respectively.  
From the Hund's third rule, 
$\langle \mbox{\boldmath$l$} \cdot \mbox{\boldmath$s$} \rangle$ is 
negative (positive) 
when the number of electrons is more (less) than half-filling.   
The experimentally observed sign change of 
the SH conductivities in the TM wires is well reproduced by 
the intrinsic SHE in the $d$-electron system.  

We now discuss the magnitude of the SH conductivities of the TM wires. 
As mentioned above, the pure spin current is absorbed perpendicularly 
into the TM wires. This means that the spin current has a distribution along 
the thickness direction, i.e., $I_{\rm S}(z)$ 
because $I_{\rm S}$ should varnish at the substrate.
The SH conductivity can be calculated by 
\begin{equation}
\sigma_{\rm SHE} =  \sigma_{\rm TM}^{2} \frac{w_{\rm TM}}{x} \left( \frac{I_{\rm C}}{\bar{I_{\rm S}}} \right) \Delta R_{\rm SHE}. \label{eq2}
\end{equation}
where $\bar{I_{\rm S}}$ is the effective spin current to contributes to ISHE. 
The factor $x$ is a correction factor 
taking into account the fact that the horizontal current driven 
by the ISHE voltage to balance the spin-orbit deflections is 
partially shunted by the Cu wire 
above the TM/Cu interface.~\cite{niimi,buhrman_prl_2011} 
The correction factor $x$ for each TM is derived from additional measurements 
of the resistance of the TM wire with and without the interface with Cu. 
It is found to be 0.36$\pm$0.08 which is not so sensitive to 
the resistivity whitin our resistivity range 
(see supplemental material in Ref.~\onlinecite{niimi}). 
We can obtain $\Delta R_{\rm SHE}$ from 
the ISHE mesurements, 
i.e., $\Delta R_{\rm SHE}=\Delta R_{\rm ISHE}$.~\cite{Kimura_SHE} 

In our case, $\lambda_{\rm TM}$ is always smaller than $t_{\rm TM}$. 
The spin currents injected into the TM wire should be diluted in the TM wire, 
which leads to a smaller $\Delta R_{\rm ISHE}$. 
To correct this effect, 
we take into account all the spin currents injected into the TM wire and then 
divide them by $t_{\rm TM}$:~\cite{niimi} 
\begin{widetext}
\begin{eqnarray}
\frac{\bar{I_{\rm S}}}{I_{\rm C}} &\equiv& 
\frac{\int_{0}^{t_{\rm TM}} I_{\rm S}(z) dz}{t_{\rm TM} I_{\rm C}} 
=\frac{\lambda_{\rm TM}}{t_{\rm TM}}\frac{\left( 1-e^{-t_{\rm TM}/\lambda_{\rm TM}} \right)^{2}}{1-e^{-2t_{\rm TM}/\lambda_{\rm TM}}} \frac{I_{\rm S}(z=0)}{I_{\rm C}} \nonumber\\
&\approx& \frac{\lambda_{\rm TM}}{t_{\rm TM}}\frac{\left( 1-e^{-t_{\rm TM}/\lambda_{\rm TM}} \right)^{2}}{1-e^{-2t_{\rm TM}/\lambda_{\rm TM}}} \frac{2 p_{\rm Py} R_{\rm Py} \sinh \left( L/2\lambda_{\rm Cu} \right) } { \left[ R_{\rm Cu} \left\{  \cosh \left( L/\lambda_{\rm Cu} \right) - 1 \right\} + 2 R_{\rm Py} \left( e^{L/\lambda_{\rm Cu}}-1 \right) \right] +2 R_{\rm TM} \sinh \left( L/\lambda_{\rm Cu} \right)}, \label{eq3}
\end{eqnarray}
\end{widetext}
where $R_{\rm Py}$ and $p_{\rm Py}$ are the spin resistance and 
the spin polarization of Py, respectively. $R_{\rm Py}$ is defined as 
$\frac{\rho_{\rm Py} \lambda_{\rm Py}}{(1-p_{\rm Py}^{2})w_{\rm Py}w_{\rm Cu}}$ 
where $\rho_{\rm Py}$ and $\lambda_{\rm Py}$ 
are the electric resistivity and 
the spin diffusion length of Py.~\cite{note_Py} 

To compare the experimentally obtained SH conductivity 
with the theoretically calculated value in Ref~\onlinecite{Kontani2}, 
we plot both of them in Fig.~4. 
In most cases, the experimental results are quantitatively consistent with 
the calculated ones within a factor of 2. 
This fact strongly suggests that the SHEs in 4$d$ and 5$d$ TMs are mainly 
caused by the intrinsic mechanism as pointed out in Ref~\onlinecite{Kontani2}.
Of course, we cannot exclude the possibility of some contributions from 
the extrinsic mechanisms such as the skew scattering and the side jump. 
However, we use a pure (at least more than 99.9\%) source for each TM 
and deposit it under a pressure of $10^{-9}$ Torr. 
This assures that no other TMs which have $d$-orbital degrees of freedom 
and cause large extrinsic SHEs are included, 
and the resistivity of the TM wire is simply caused by grain boundary, 
lattice mismatch, other defects and so on. 
We believe that the contribution from the intrinsic mechanism 
is more dominant in 4$d$ and 5$d$ TMs than that from the extrinsic one. 

In the previous works on the SHE in Pt reported 
by some of the present authors,~\cite{Kimura_SHE,Vila_SHE} 
the mechanism of the SHE was the extrinsic 
one (side-jump scattering) and the SH angle ($\alpha_{\rm H}$) 
of Pt was 0.37\%. 
In the present study, however, we claim that 
the dominant mechanism of the SHE in Pt is intrinsic one and 
the SH angle for Pt is 2.1\%, which is about 6 times larger. 
There are several reasons for our present conclusions; 
in the previous work,~\cite{Vila_SHE} they concluded that 
the side-jump mechanism was dominant because the SH resistivity 
is proportional to $\rho_{\rm Pt}^{2}$.
However, this resistivity dependence is also predicted in the intrinsic 
mechanism.~\cite{Kontani1,Kontani2,Kontani3} 
For the SH angle of Pt, the boundary condition 
for $\bar{I_{\rm S}}/I_{\rm C}$ had not been taken into account 
appropriately in the previous study 
(see Eq.~(2) in Ref.~\onlinecite{Vila_SHE}). 
In the present study, on the other hand, we impose $I_{\rm S} = 0$ 
at the bottom of the TM wires. 
In addition, we consider the shunting effect of the Cu strip 
as detailed in Ref.~\onlinecite{niimi}. 
The resistivity of the TM wires is much larger than that of the Cu strip 
($\rho_{\rm Cu} = 1.5$ $\mu \Omega \cdot$cm). 
This causes a smaller current flowing the TM wires and as a result causes the 
underestimation of the SH angle 
as discussed in Refs.~\onlinecite{niimi} and~\onlinecite{buhrman_prl_2011}. 
The SH angle (2.1\%) of Pt in the present paper 
is consistent with that (1.3\%) in Ref.~\onlinecite{Hoffmann2} 
and is a few times 
smaller than that (5.6\%) in Ref.~\onlinecite{buhrman_prl_2011}.


\section{Conclusion}

In conclusion, we have measured the SH conductivities for 
various TMs in a lateral spin valve structure. 
When $d$ electrons are smaller (larger) than the half-filled value, 
we have observed negative (positive) SH conductivies. 
Compared to the recent theoretical calculations based on 
the intrinsic SHEs in 4$d$ and 5$d$ TMs, the experimentally obtained 
SH conducitivies are semiquantitativley consistent with the theoretical ones. 
This fact strongly indicates that the intrinsic mechanism based on 
the degeneracy of $d$-orbits in the LS coupling is dominant for the SHEs 
in 4$d$ and 5$d$ TMs.

\section{Acknowledgements}

We acknowledge helpful discussions with J. Inoue, 
N. Nagaosa, S. Takahashi, and S. Maekawa. 
We would also like to thank Y. Iye and S. Katsumoto 
for the use of the lithography facilities. 
This work was supported by a
Grant-in-Aid for Scientific Research in Priority Area ``Creation
and Control of Spin Current" (Grant No. 19048013)
from the Ministry of Education, Culture, Sports, Science and
Technology of Japan.


\begin{thebibliography}{00}
\bibitem{Maekawa} 
S. Maekawa, Nature Mater. {\bf 8}, 777 (2009). 
\bibitem{SHE1} 
M. I. Dyakonov, Int. J. Mod. Phys. B, {\bf 23}, 2556 (2009).
\bibitem{SHE2} 
J. E. Hirsch, Phy. Rev. Lett. {\bf 83}, 1834 (1999). 
\bibitem{SHE3} 
S. Zhang, Phys. Rev. Lett. {\bf 85}, 393 (2000).
\bibitem{Kimura_SHE} 
T. Kimura, Y. Otani, T. Sato, S. Takahashi, and S. Maekawa, Phys. Rev. Lett. {\bf 98}, 156601 (2007).
\bibitem{Vila_SHE} 
L. Vila, T. Kimura, and Y. Otani, Phys. Rev. Lett. {\bf 99}, 226604 (2007).
\bibitem{Extrinsic1} 
J. Smit, Physica (Amsterdam) {\bf 24}, 39 (1958). 
\bibitem{Extrinsic2} 
L. Berger, Phys. Rev. B {\bf 2}, 4559 (1970).
\bibitem{Extrinsic3} 
A. Fert and O. Jaoul, Phys. Rev. Lett. {\bf 28}, 303 (1972).
\bibitem{Intrinsic1} 
R. Karplus and J. M. Luttinger, Phys. Rev. {\bf 95}, 1154 (1954).
\bibitem{Intrinsic2} 
N. Nagaosa, J. Phys. Soc. Jpn. {\bf 75}, 042001 (2006). 
\bibitem{murakami}
S. Murakami, N. Nagaosa, S.-C. Zhang, Science {\bf 301}, 1348 (2003). 
\bibitem{sinova}
J. Sinova, D. Culcer, Q. Niu, N. A. Sinitsyn, T. Jungwirth, and A. H. MacDonald, Phys. Rev. Lett. {\bf 92}, 126603 (2004). 
\bibitem{Lee} 
W.-L. Lee, S. Watauchi, V. L. Miller, R. J. Cava, and N. P. Ong, Science {\bf 303}, 1647 (2004).
\bibitem{Miyasako} 
T. Miyasato, N. Abe, T. Fujii, A. Asamitsu, S. Onoda, Y. Onose, N. Nagaosa, and Y. Tokura, Phys. Rev. Lett. {\bf 99}, 086602 (2007). 
\bibitem{Jin} 
Y. Tian, L. Ye, and X.-F. Jin, Phys. Rev. Lett. {\bf 103}, 087206 (2009). 
\bibitem{AHE1} 
S. Onoda, N. Sugimoto and N. Nagaosa, Phys. Rev. Lett. {\bf 97}, 126602 (2006).
\bibitem{note_intrinsic}
In Ref.~\onlinecite{AHE1}, the extrinsic term is more dominant in the clean regime than the intrinsic one, which seems to be inconsistent with Refs.~\onlinecite{murakami} and~\onlinecite{sinova}. However, the extrinsic term strongly depends on the property of impurtiy, in other words, its orbital degrees of freedom as detailed in Ref.~\onlinecite{Kontani1}. In this sense, the statement about the extrinsic term in Ref.~\onlinecite{AHE1} is not general. 
\bibitem{Guo} 
G. Y. Guo, S. Murakami, T.-W. Chen, and N. Nagaosa, Phys. Rev. Lett. {\bf 100}, 096401 (2008).
\bibitem{Kontani1} 
H. Kontani, M. Naito, D. S. Hirashima, K. Yamada, and J. Inoue, J. Phys. Soc. Jpn. {\bf 76}, 103702 (2007).
\bibitem{Kontani2} 
T. Tanaka, H. Kontani, M. Naito, T. Naito, D. S. Hirashima, K. Yamada and J. Inoue, Phys. Rev. B {\bf 77}, 165117 (2008).
\bibitem{Kontani3} 
H. Kontani, T. Tanaka, D. S. Hirashima, K. Yamada, and J. Inoue, Phys. Rev. Lett. {\bf 102}, 016601 (2009).
\bibitem{review_otani}
See, for example, Y. Otani and T. Kimura, Physica E {\bf 43} 735 (2011).
\bibitem{Takahashi} 
S. Takahashi and S. Maekawa, Phys. Rev. B {\bf 67}, 052409 (2003).
\bibitem{notation} 
Equation~(\ref{eq1}) in the present paper is different from Eq.~(\ref{eq1}) in Ref.~\onlinecite{Vila_SHE} by a factor of 2 in front of the hyperbolic sine term. In the present manuscript we take into account only one path for spin relaxation whose direction is perpendicular to the plane of the TM middle wires. On the other hand, in Ref.~\onlinecite{Vila_SHE} there are two paths for the spin relaxation, i.e., side edges of the TM wires. Since the spin diffusion lengths of the TM wires are extremely small, spin current should be absorbed perpendicularly into the plane of the TM wires, not from the side edges. It is known that the former model gives a better fitting for the distance dependence of NLSV than the latter one (see Ref.~\onlinecite{Kimura1}). 
\bibitem{note_spin_resistance}
The spin resistance defined here is expressed by $\rho \lambda/A$. This means that a larger spin resistance corresponds to less spin diffusion process, which is intuitively opposite to a normal resitance. 
\bibitem{niimi}
Y. Niimi, M. Morota, D. H. Wei, C. Deranlot, M. Basletic, A. Hamzic, A. Fert, and Y. Otani, Phys. Rev. Lett. {\bf 106}, 126601 (2011).
\bibitem{Kimura1} 
T. Kimura, T. Sato and Y. Otani, Phys. Rev. Lett. {\bf 100}, 066602 (2008).
\bibitem{Hoffmann1} 
O. Mosendz, J. E. Pearson, F. Y. Fradin, G. E. W. Bauer, S. D. Bader and A. Hoffmann, Phys. Rev. Lett. {\bf 104}, 046601 (2010). 
\bibitem{Hoffmann2}
O. Mosendz, V. Vlaminck, J. E. Pearson, F. Y. Fradin, G. E. W. Bauer, S. D. Bader, and A. Hoffmann, Phys. Rev. B {\bf 82}, 214403 (2010). 
\bibitem{buhrman_prl_2011}
L. Liu, T. Moriyama, D. C. Ralph, and R. A. Buhrman, Phys. Rev. Lett. {\bf 106}, 036601 (2011)
\bibitem{note_Py}
$\rho_{\rm Py}$, $p_{\rm Py}$ $\lambda_{\rm Py}$ are 19 $\mu \Omega \cdot$cm, 0.23 and 5 nm at 10 K, respectively.
\end{thebibliography}
\end{document}